\documentclass[twocolumn,showpacs,preprintnumbers,amsmath,amssymb]{revtex4}

\usepackage{graphicx}
\usepackage{dcolumn}
\usepackage{bm}

\begin{document}

\preprint{APS/123-QED}

\title{Theoretical study of one-proton removal from $^{15}$O.}

\author{Y. L. Parfenova}
\email{Yulia.Parfenova@ulb.ac.be}
\affiliation{%
Physique Nucl\'eaire Th\'eorique et
Physique Math\'ematique, CP229, Universit\'e Libre de Bruxelles
B 1050 Brussels, Belgium.}
\altaffiliation[Also at ]{ Skobeltsyn Institute of Nuclear Physics, Moscow State
University, 119992 Moscow, Russia}
\author{M.V. Zhukov}%
\affiliation{%
Fundamental Physics, Chalmers University of Technology,
S-41296 G\"{o}teborg, Sweden\\
}%


\date{\today}

\begin{abstract}
One-proton removal from $^{15}$O at intermediate energies (56 A
MeV) is studied in the eikonal approximation of the Glauber model.
The production of the $^{14}$N core fragment in the ground and
excited states is regarded. The calculated proton removal cross
section, the $^{15}$O interaction cross section, and the
longitudinal momentum distribution of the $^{14}$N fragments are
compared to recent experimental data \cite{Jep04}.
\end{abstract}

\maketitle

\section{Introduction \label{secI}}

During the last decade two-proton emitters and proton-rich nuclei,
in the vicinity of the proton drip-line, are subjects of intensive
experimental and theoretical studies. In particular the study of
a candidate to possess a two-proton halo in the ground state,
namely the Borromean $^{17}$Ne nucleus is of special experimental
and theoretical interest (see, for example, discussions in
\cite{zhu95,kan03,Gri05}). The inherent feature of the halo
structure is a relatively small separation energy of a valence
nucleon. It reveals itself in a large valence nucleon removal
cross section and a narrow core longitudinal momentum (LM)
distribution.

In the case of $^{17}$Ne the proton removal cross section,
measured at the energy 66 A MeV on a Be target \cite{kan03}, is
relatively large compared to the cluster model ($^{15}$O$+p+p$)
predictions \cite{Gri05}. At the same time, the measured core LM
distribution is wider than the calculated one. Both these facts
can be attributed to contribution of a proton removal from the
$^{15}$O core in $^{17}$Ne if this cross section is relatively
large.


Recently the $^{14}$N longitudinal momentum (LM) distribution and
break-up cross section (into the $^{14}$N+p channel) have been
measured in fragmentation of $^{15}$O on a Be target at the energy
56 A MeV \cite{Jep04}. This opens a possibility for more precise
calculations of the proton removal from $^{17}$Ne and evaluations
of the contribution of the proton removal from the $^{15}$O core
to this process.

In this paper, we present a detailed analysis of the $^{15}$O
break-up in light targets. We perform the calculations in the
eikonal approximation of the Glauber model
\cite{Hen96,Esbens1,Esb00,Par00}. This approach is well developed
and convenient for calculations of break-up cross sections,
interaction cross sections, and momentum distributions of
fragments in break-up of a nucleus at intermediate and high
energies (from 30 to 1000 A MeV).

The formalism for the calculations is described in Section
\ref{secII}. The main ingredients of the Glauber model are the
wave function of the relative motion of the fragments and the
profile functions defining the fragment-target interaction. They
are fixed using experimental data on the nucleon-nucleus and
nucleus-nucleus cross sections, proton separation energies, the
level scheme of the core nucleus, etc. In particular, the profile
functions are fitted using the nucleus-nucleus and nucleon-nucleus
interaction cross sections.

The wave function is obtained in the core+proton
($^{14}$N+$p$) model of $^{15}$O, where the $^{14}$N
core fragment can be in the ground and excited states
(see, for example, Ref. \cite{Par00}).
The $p$-wave proton removal from $^{15}$O ($J^{\pi}=1/2^-$)
leads to few $^{14}$N bound states. We consider four of them (for
details see \cite{Sne69}): $E_x=0.0$ MeV ($J^{\pi}=1^+$, $T=0$),
$E_x=2.313$ MeV ($0^+$, $1$), $E_x=3.948$ MeV ($1^+$, $0$), and
$E_x=7.029$ MeV ($2^+$, $0$). Here, $E_x$ is the excitation energy
and ($J^{\pi}$, $T$) are the spin and isospin of the $^{14}$N
state. For each state, the depth of the ($^{14}$N+$p$) interaction
potential (see, below) is fitted to reproduce the proton
separation energy.

The cross sections of the proton removal from
the $^{15}$O ground state
are determined by the spectroscopic factors \cite{Sne69,Coh67}
of the $p$-wave proton states.


In Section \ref{secIII} we fit the profile functions in
calculations of the corresponding nucleus-nucleus and
proton-nucleus interaction cross sections, and compare results to
the available experimental data.

In Section \ref{secIV} we present the calculated cross sections
and longitudinal momentum distributions of the $^{14}$N fragments
produced in various states in the process of the one-proton
removal from $^{15}$O on a $^{9}$Be target. These results are
compared to the experimental data on the $^{14}$N longitudinal
momentum distribution and the break-up cross section measured at
the energy 56 A MeV \cite{Jep04}.

\section{Cross sections and momentum distributions \label{secII}}
In the core-nucleon model of the projectile nucleus, the initial
state is described by the wave function (WF) $\Psi
_{JM_J}(\vec{r})$ of the core-nucleon relative motion with a total
angular momentum $J$ and its projection $M_J$. The WF depends on
the relative coordinate $\vec{r}$ between nucleon and core.

After
interaction with a target, the WF of the projectile will be
corrected by factors, connected with nucleon-target and
core-target interactions. Thus, the WF in the projectile rest
frame is modified as \cite{Hen96}
\begin{equation}
\Psi (\vec{r},\vec{R})= \,S_{c}(b_{c})S_{n}(b_{n})\Psi _{JM_J}(\vec{r}),  \label{WF}
\end{equation}
where $\vec{R}$ is the coordinate of the center of mass of
the projectile, $b_{i}=|\vec{b}_{i}|$ ($i=n,c$),
and $\vec{b}_{n}$, $\vec{b}_{c}$ are the transverse two
dimensional impact parameters of the nucleon
and the core with respect to the target nucleus, i.e.
$\vec{b}_{n}=\vec{R}_{\perp }+\vec{r}_{\perp }A_{c}/(A_{c}+1)$
and $\vec{b}_{c}=\vec{R}_{\perp }-\vec{r}_{\perp }/(A_{c}+1)$,
where $\vec{R}_{\perp }$ and $\vec{r}_{\perp }$
are components, perpendicular to the beam direction taken
as $z$ axis, and $A_{c}$ is the mass number of the core.
The profile functions $S_{n}(b_{n})$ and $S_{c}(b_{c})$ are
generated by nucleon and core interactions with the
target nucleus. 

The fragmentation includes nucleon stripping and
diffraction processes.
The corresponding cross sections are given by the
equations \cite{Hen96}
\begin{widetext} 
\begin{eqnarray}
\sigma _{str} &=&{\frac{1}{2L+1}}\sum_{M}\,\int d\vec{R}_{\perp
}\;\int d\;\vec{r}\;\Psi _{LM}^{\ast }(\vec{r})\;(\,1-|S_{n}|^{2})%
\;|S_{c}|^{2}\;\Psi _{LM}(\vec{r})  \nonumber \\
\sigma _{diff} &=& {\frac{1}{2L+1}}\sum_{M}\, \int d\vec{R}%
_{\perp }\;\int d\;\vec{r}\;\Psi _{LM}^{\ast }(\vec{r}%
)\;|S_{n}S_{c}|^{2}\;\Psi _{LM}(\vec{r})   \label{CrSec} \\
&&-\frac{1}{2J+1} \sum_{MM^{\prime }}\int d\vec{R}_{\perp }\;\left| \int d\;\vec{r}\;\Psi
_{JM_J^{\prime }}^{\ast }(\vec{r})\;S_{n}S_{c}\;\Psi _{JM_J}(\vec{r}%
)\right| ^{2}   . \nonumber
\end{eqnarray}
\end{widetext} 

The proton removal cross section is found as the
sum $\sigma _{-p}=\sigma _{str}+\sigma _{diff}$ (\ref{CrSec}).

The wave function $\Psi _{JM_J}$ is
\begin{equation}
\Psi _{JM_J}=\left[ \left[ \Psi _{LM}(\vec{r}) \otimes
\chi_{s_n m_n} \right]_{j_n} \otimes \chi_{s_c m_c} \right]_{JM_J} ,
\end{equation}
where $\chi_{s_c m_c}$ is the internal wave function of
the core including the spin function, $\chi_{s_n m_n}$ is
the spin function of the valence nucleon.

We denote the part of the WF related to the relative motion as
$\Psi _{LM}$
\begin{equation}
 \Psi _{LM}(\vec{r}) = R_{L}(r) Y_{LM},
\end{equation}
where $Y_{LM_L}$ is the spherical function.

The radial part of the core-proton WF, $R_{L}(r)$, is obtained as
a solution of the Schr\"{o}dinger equation for the Woods-Saxon
potential (the Coulomb $^{14}N+p$ potential is also included). For
each state of $^{14}$N, the parameter $V_{0}$ of the Woods-Saxon
potential is fitted to reproduce the proton separation energy with
the fixed parameters $a_{0}=0.65$ fm and $R_{0}=1.25 A^{1/3}$=3.00 fm.
The depth parameters and the proton separation energies are given
in Table \ref{Tab1}.

In the calculations of the cross sections and LM distributions of
the fragments we consider the $p$ ($p_{1/2}$ and $p_{3/2}$) proton
removal. The $p$-wave proton removal from $^{15}$O leads to the
residual $^{14}$N core in the bound states $E_x=0.0$ MeV
($J^{\pi}=1^+$, $T=0$), $E_x=2.313$ MeV ($0^+$, $1$) $E_x=3.948$
MeV ($1^+$, $0$), and $E_x=7.029$ MeV ($2^+$, $0$).

Note, that the spectroscopic factors are not measured yet for
$^{15}$O.
As was shown in the DWBA analysis \cite{Bom71},
the contribution of the
protons with $l=1$ dominates
in the proton transfer reaction leading to the ground state of
$^{15}$O.
The spectroscopic factors of the states can be taken
as those predicted by Cohen and Kurath \cite{Coh67}.
These values are close to the measured values in the neutron
pickup reactions with the mirror $^{15}$N nucleus, the
$^{15}$N$(p,d)^{14}$N reaction with 40 MeV protons \cite{Sne69},
and the $^{15}$N$(d,t)^{14}$N reaction with 90 MeV deuterons
\cite{Sah89} (see the discussion in \cite{Jep04} and references
therein). We use the spectroscopic factors from \cite{Coh67} and
\cite{Sah89}. These factors $C^2S$ are also listed in Table
\ref{Tab1}.

\begin{table}[tbp]
\caption{\label{Tab1} The depth parameter $V_{0}$ of the
Woods-Saxon potential, obtained with the diffuseness parameter
$a_{0}=0.65$ fm and radius $R_0=3.00$ fm for the $p$-wave proton
separation energy $E_{s}$. $E_{x}$ is the corresponding $^{14}$N
core excitation energy. $C^{2}S$ are the spectroscopic factors,
$^a$ \cite{Coh67} and $^b$ \cite{Sah89}.}
\begin{ruledtabular}
\begin{tabular}{cccccc}
      &               &          &          & \multicolumn{2}{c}{Woods-Saxon potential}  \\
$E_{x}$ &   $^{14}$N  & $C^{2}S^a$ & $C^{2}S^b$ & $V_{0}$  & $E_{s}$  \\
MeV   & $(J^{\pi},T)$ &          &          & (MeV)    & (MeV)    \\ \hline
0     & $(1^{+},0)$   & $1.459$  & $1.343$  & $-48.09$ & 7.297    \\
2.313 & $(0^{+},1)$   & $0.418$  & $0.472$  & $-52.07$ & 9.610    \\
3.948 & $(1^{+},0)$   & $0.696$  & $0.656$  & $-54.78$ & 11.245   \\
7.029 & $(2^{+},0)$   & $1.250$  & $1.250$  & $-59.73$ & 14.326   \\
\end{tabular}
\end{ruledtabular}
\end{table}


Note, that the contribution of the  $^{14}$N excited bound states
to the diffraction cross section (\ref{CrSec}) is relatively small
and is neglected here.

The LM distributions of the core fragments are obtained by the
Fourier transformation of the core-proton WF, $R_{L}(r)$,
corrected for the core-target and nucleon-target interactions
\begin{eqnarray}
{\frac{d\sigma _{str}}{dk_{z}}} &=&
\frac{1}{2L+1} \int \limits_0^{\infty} b_{n} db_{n}
(1-|S_{n}(b_{n})|^{2}) \nonumber \\
& & \int\limits_{0}^{\infty } r_{\perp }dr_{\perp} d \phi
|S_{c}(|\vec{b}_{n}-\vec{r}_{\perp }|)|^{2}
\label{distR} \\
& & \sum \limits_{M_L} \left| \int \limits_{-\infty }^{\infty}
e^{ik_{z}z}R_{L}\;(\sqrt{r_{\perp }^{2}+z^{2}})
Y_{LM_L} dz \right| ^2 . \nonumber
\end{eqnarray}

The core longitudinal momentum distribution in the diffraction
breakup is assumed to be similar \cite{Hen96,Neg99}
to that of stripping.

The expression (\ref{distR}) gives the contribution of the
LM distribution coming from each neutron-core state
composing the $^{15}$O ground
state wave function. For comparison with the experimental
data, we sum up these contributions weighted by the spectroscopic
factors (Table \ref{Tab1}).

The fragment-target interaction cross section 
is determined by the profile function $S_{\nu }(\vec{b}_{\nu })$ as
(\ref{profile}) as 
\begin{equation}
\sigma _{I}^{\nu }=\int d^{2}\vec{b}_{\nu }\,(\,1-|S_{\nu }(\vec{b}_{\nu
})|^{2}) ,  \label{FragRecCS}
\end{equation}
where index
$\nu $ denotes the fragment ($\nu =p$,$^{14}$N), $\vec{b}_{\nu }$
is the impact parameter of the $\nu$-th fragment.

The interaction cross section for the fragmented projectile is
expressed through the profile functions of the fragment-target
interaction and the wave function of the relative motion of the
fragments.
\begin{eqnarray}
\sigma _{I} &=& \frac{1}{2J+1}
\sum_{MM^{\prime }}\int d\vec{R}_{\perp }   \label{IntSec} \\
&&\left[ 1-\left| \int d\;\vec{r}\;\Psi
_{JM_J^{\prime }}^{\ast }(\vec{r})\;S_{n}S_{c}\;\Psi _{JM_J}(\vec{r}%
)\right| ^{2} \right] .  \nonumber
\end{eqnarray}

\section{Profile functions
\label{secIII}}

The profile function of the fragment-target interaction in
(\ref{WF}),(\ref{CrSec}), and (\ref{distR})
is determined as an integral
of the corresponding complex interaction potential
\begin{equation}
S_{\nu }(\vec{b}_{\nu })=\exp \left[ -\frac{i}{\hbar v}\int\limits_{-\infty
}^{\infty }dz\;\,V_{\nu T}\left( \sqrt{b_{\nu }^{2}+z^{2}}\right) \right] ,
\label{profile}
\end{equation}
where $V_{\nu T}(r)$ is the fragment-target interaction potential,
$v$ is the $^{15}$O beam velocity in the laboratory frame. The
fragment-target interaction potential is determined by folding of
the fragment density distribution and the nucleon-target
interaction potential.

To calculate the nucleon-target interaction potential $V_{\nu
T}(r)$ ($\nu =n$, $p$) at energies less than 65 A MeV, we use the
parameters of the global nucleon-nucleus optical potential
\cite{OP1}. We also use the interaction potential \cite{Hen96}
generated from the free nucleon-nucleon ($NN$) interaction
\cite{NN1,NN2} valid at energies from 10 to 2000 A MeV. In this
case, the nucleon-target interaction potential is obtained by
folding of the target density distribution and the nucleon-nucleon
interaction potential.
For the details of the profile function calculations we
refer to \cite{Par00,Par02}.



For description of the target and fragment nuclear densities we
use different parametrizations. The $^9$Be and $^{14}$N densities
are parameterized in the harmonic oscillator model \cite{deV87}
\begin{equation}
\rho (r)=\rho _{0}[1+\alpha (r/a)^{2}]\exp (-(r/a)^{2})\;.  \label{denHO}
\end{equation}
The parameter $\alpha$ is related to $a$ \cite{deV87}.
The parameter $\alpha$ is fitted (see the next Section) to
reproduce nucleon-nucleus and nucleus-nucleus interaction
cross sections.


The $^{12}$C density distribution
is approximated by a sum of Gaussians \cite{deV87} as
\begin{equation}
\rho (r)=\textstyle\sum \limits_{i} A_{i}\left(
e^{- (r-\beta R_{i})^2/\gamma^2}
+e^{- (r+\beta R_{i})^2/\gamma^2}\right)  \label{denGAF}
\end{equation}
with the parameters from Ref. \cite{deV87}. In order to vary the
calculated cross section obtained with the density distribution
(\ref{denGAF}),
we introduce a scaling factor
$\beta$ and replace $R_i$ by $\beta R_i$ in (\ref{denGAF}).

All the distributions $\rho $ are normalized to unity, and
$\rho _{0}$ is a normalization factor.




To fit the profile functions, corresponding experimental data for
interaction (reaction) cross sections on C and Be targets at
intermediate and high energies are used.


In the case of $^{12}$C, the scaling parameter $\beta$ in
(\ref{denGAF}) is fitted to reproduce the experimental data for
$^{12}$C+$^{12}$C
\cite{Oza01,Sah86,She89,Wei03,Tan85,Kox85,Kox87,Jar87,Tan90,Bar93} and
$p+^{12}$C \cite{Bar93,NNDC} interaction cross sections. The best
fit is achieved for $\beta$=0.94. With this $\beta$ value the
$^{12}$C rms radius is 2.37 fm, that is close to the $^{12}$C rms
matter radius 2.33 fm
obtained in \cite{Oza01}.

\begin{figure}
\includegraphics{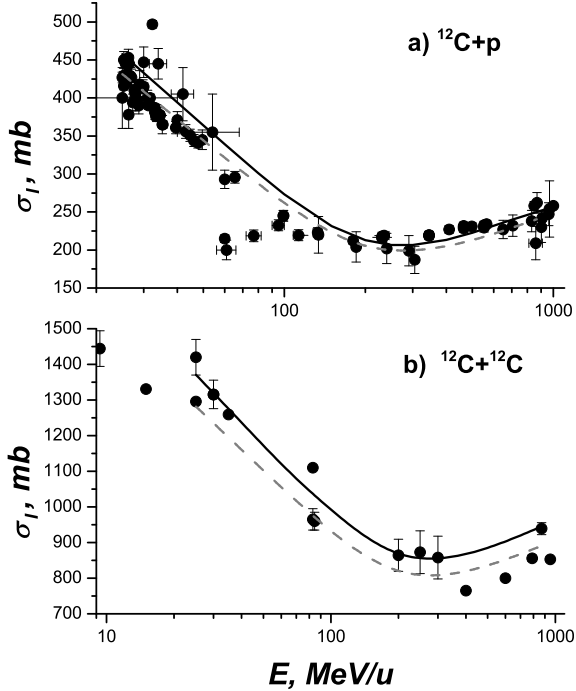}
\caption{\label{Fig1} The energy dependence of the $p$+$^{12}$C
and $^{12}$C+$^{12}$C interaction cross sections, $\sigma_I$,
calculated with the NN interaction potential. Dots in a) and b)
are the experimental data  \cite{Bar93,NNDC} and
\cite{Oza01,Sah86,She89,Wei03,Tan85,Kox85,Kox87,Jar87,Tan90,Bar93},
respectively. The curves correspond to $\beta$=0.94 (dashed gray
lines) and $\beta$=1 (solid black lines). }
\end{figure}

Figures \ref{Fig1}a,1b show the calculated (dashed gray curves)
and measured (dots)
$p+^{12}$C and
$^{12}$C$+^{12}$C interaction cross sections at energies from 20
to 1000 A MeV. For comparison, the cross sections obtained with
the charge radius of carbon $r_c=2.47$ fm ($\beta$=1) are also given
in Fig. \ref{Fig1} (solid black curves).

The $^{9}$Be density parameter $a$ in (\ref{denHO}) is fitted to
reproduce the experimental data on the $p+^{9}$Be
\cite{Bar93,NNDC} and $^9$Be$+^9$Be \cite{Tan85} interaction
cross sections. The value $a=1.69$ fm corresponds the $^{9}$Be rms
radius 2.38 fm \cite{Oza01}.
To have a measure of sensitivity of the results to the input
parameters of the model, we present the results of the
calculations with the parameter $a=1.79$ fm, also allowing a good
fit of the $^9$Be-nucleus cross section.

\begin{figure}
\includegraphics{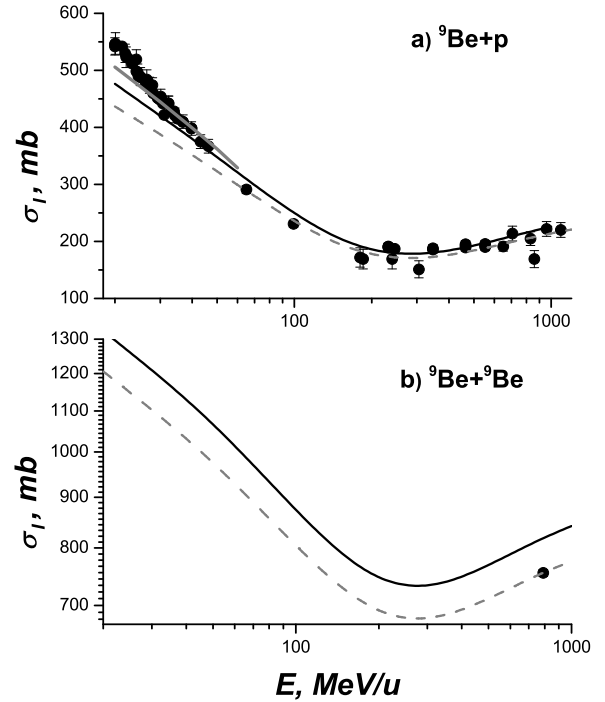}
\caption{\label{Fig2} The energy dependence of the $p$+$^{9}$Be
and $^{9}$Be+$^{9}$Be interaction cross sections, $\sigma_I$. Dots
in a) and b) are the experimental data \cite{Bar93,NNDC} and
\cite{Tan85}, respectively. Dashed gray and
solid black lines are the calculations with the
NN interaction potential with the parameters $a=1.69$ and
$a=1.79$ fm. The solid grey line are the calculations with the 
optical model potential (OMP).}
\end{figure}

In Figures \ref{Fig2}a and \ref{Fig2}b the calculated cross
sections are compared to the experimental data. These results are
also compared to the calculations with the Be rms radius equal to
the Be charge radius, 2.52 fm ($a=1.79$ fm) \cite{deV87}.

The $p+^{9}$Be interaction cross section calculated with the
NN-interaction potential at energies less than 60 A MeV is
underestimated, while that obtained with the optical model
potential satisfy the experimental data. At higher energies, the
cross section calculated with the NN-interaction potential is in a
good agreement with the experimental data \cite{Bar93}.

To test the fitted density parameters of the $^{12}$C and $^9$Be
we calculate the interaction cross section in the
$^{9}$Be+$^{12}$C reaction at the energy 790 A MeV. The value,
818.7 mb, is very close to the experimental one 806(9) mb
\cite{Oza01}.

Using the experimental data on the $^{14}$N+p reaction
\cite{Bar93} and the $^{14}$N+$^{12}$C reaction at the energies
39.3 \cite{Fan00} and 965 A MeV \cite{Oza01} we found $a=1.76$ fm
in the $^{14}$N density parametrization (\ref{denHO}). This value
corresponds to the $^{14}$N rms matter radius 2.44 fm known from
experiment.

\begin{figure}
\includegraphics{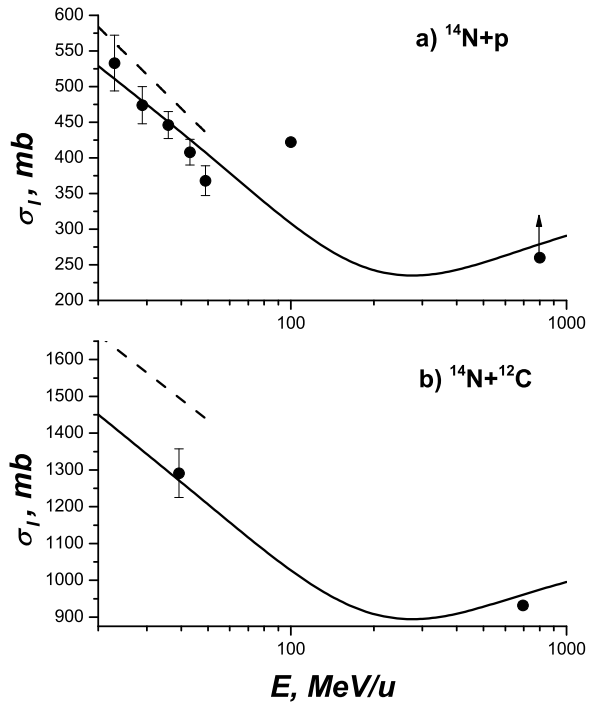}
\caption{\label{Fig3} The energy dependence of the $p$+$^{14}$N
and $^{14}$N+$^{12}$C interaction cross sections, $\sigma_I$. Dots
in a) and b) are the experimental data \cite{Bar93,NNDC} and
\cite{Oza01,Fan00}, respectively. The calculations with the NN
interaction potential and the optical potential (OMP) are shown by
solid and dashed lines, respectively.}
\end{figure}

The results of these calculations and the experimental data are
given in Fig. \ref{Fig3}. The cross sections calculated with the
NN interaction potential are in a better agreement with the
experimental data both for the proton-nucleus and the
nucleus-nucleus interaction cross sections than those obtained
with the optical model potential.

For further calculations of the $^{15}$O break-up on a Be target
at the energy 56 A MeV
we use profile functions obtained with the NN interaction
potential.

With the $^{14}$N rms radius we can estimate
the $^{15}$O rms radius as
\begin{equation}
r_{m}^2(^{15}{\text O})=\frac{A_{c}A_p}{A^2} \left \langle
r_{c-p}^2 \right \rangle + \frac{A_c}{A}r_{m}^2(^{14}{\text N}) ,
\end{equation}
where $r_{c-p}$ is the distance of the valence proton from the
$^{14}$N center of mass, $A=A_c+A_p$ is the mass number of the
projectile, the valence proton mass number $A_p=1$.

With the $^{14}$N rms matter radius $r_{m}=2.44$ fm, which
corresponds ($r_{c}^2=r_{m}^2+0.8^2$) to the charge radius
$r_{c}(^{14}$N$)= 2.57$ fm \cite{deV87}, and the rms $r_{c-p}$
distance of the proton $\left \langle r_{c-p}^2 \right
\rangle^{\frac{1}{2}}=3.15$ fm, the $^{15}$O rms matter radius is
$r_{m}(^{15}$O$)= 2.48$ fm. This value is consistent with the
values obtained in Refs. \cite{Oza01,War05}. The corresponding
$^{15}$O rms charge radius is $r_{c}(^{15}$O$)=2.61$ fm.

\begin{table}[tbp]
\caption{\label{Tab2} The calculated ($\sigma_I$)
and measured ($\sigma_I^{exp}$) nucleus-nucleus
interaction cross sections.}
\begin{ruledtabular}
\begin{tabular}{ccccc}
Proj.    & Target   & E       & $\sigma_I$ & $\sigma_I^{exp}$  \\
         &          & (MeV/u) & (mb)       & (mb)              \\  \hline
$^{15}$O & $^{9}$Be & 710  &  881$^{a}$ &  912(23)    \\
         & $^{9}$Be & 710  &  920$^{b}$ &             \\
         & $^{12}$C & 670  &  939       &  915(13)    \\
         & $^{12}$C & 710  &  945       &  922(49)    \\
\end{tabular}
\end{ruledtabular}
\\$^{a}$ obtained with $a=1.694$ fm ~~~~~~~~~~~~~~~~~~~~~~~~~~~~~~~~~~~~~~~~~~~~~
\\$^{b}$ obtained with $a=1.791$ fm ~~~~~~~~~~~~~~~~~~~~~~~~~~~~~~~~~~~~~~~~~~~~~
\end{table}

In Table \ref{Tab2}, the values of the $^{15}$O
interaction cross section (\ref{IntSec})
obtained in the $^{12}$C and $^9$Be
targets with the fitted density parameters are compared
to the experimental data. One can see
a good agreement with the experimental data \cite{Oza01}.

\section{Results and discussion \label{secIV}}

\begin{table}[tbp]
\caption{\label{Tab3} The single-particle one-proton removal cross
section ($\sigma_{-p}^{sp}$) and the one-proton removal cross
section ($\sigma_{-p}$) from $^{15}$O calculated at the energy 56
A MeV on a Be target.}
\begin{ruledtabular}
\begin{tabular}{cccc}
$^{14}$N $(J^{\pi},T)$ &  $\sigma_{-p}^{sp}$  &  $\sigma_{-p}$ &  FWHM  \\
                      &   (mb)          &   (mb)          &  (MeV/c) \\ \hline
 $(1^+,0)$            &  29.7           &  43.3           &  178    \\        
 $(0^+,1)$            &  25.9           &  10.8           &  191    \\        %
 $(1^+,0)$            &  23.3           &  16.2           &  198    \\        
 $(2^+,0)$            &  20.4           &  25.6           &  209    \\ \hline 
Total                 &                 &  95.9           &  191   \\ 
\end{tabular}
\end{ruledtabular}
\end{table}


The $^{14}$N and $^{15}$O interaction cross sections obtained at
the energy 56 A MeV on a Be target are $\sigma_I(^{14}N)=1061$ mb
and $\sigma_I(^{15}O)=1091$ mb, respectively.

One-proton removal cross sections from $^{15}$O
and the corresponding FWHM values of the LM distribution of the
$^{14}$N fragments obtained at the energy 56 A MeV for a Be
target, are listed in Table \ref{Tab3}. All the values are
calculated with the Be target density parameter $a=1.69$ fm. The
single particle proton removal cross sections, $\sigma_{-p}^{sp}$,
and those multiplied by the corresponding spectroscopic factors \cite{Coh67},
$\sigma_{-p}$, are given for each single particle state.

The total value of the one-proton removal cross section and the LM
distribution (last row of Table \ref{Tab3}) are found as the sum
of the proton removal cross sections $\sigma_{-p}$ and the
corresponding LM distributions.

\begin{figure}
\includegraphics{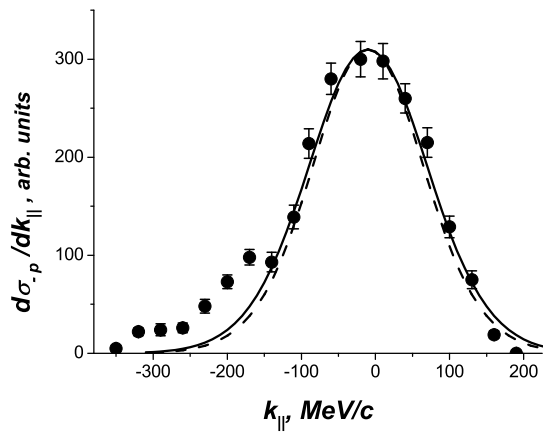}
\caption{\label{Fig5} Total longitudinal momentum distribution of
the $^{14}$N fragments (solid line) from the $^{15}$O break-up on
Be target at the energy 56 A MeV. Dots are the experimental data
\cite{Jep04}. Dashed line shows the longitudinal momentum
distribution of the $^{14}$N in the ground state. }
\end{figure}

The calculated values of the total proton-removal cross section
and the FWHM (Table \ref{Tab3}) obtained with the Be target
density parameter $a=1.69$ fm and the spectroscopic factors
\cite{Coh67} are in a very good agreement with the experimental
values $80 \pm 20$ mb and $190 \pm 10$ MeV/c  \cite{Jep04}. With
the spectroscopic factors from \cite{Sah89} the value of the
proton removal cross section is 92.0 mb. With larger target
density parameter $a=1.79$ fm we get a larger values of the cross
sections. In this case, the total proton removal cross section
obtained with the spectroscopic factors \cite{Coh67} is
$\sigma_{-p}=100.2$ mb. So one can see that the total one proton
removal cross section is not very sensitive to small variations of
spectroscopic factors or target density parameter.

In Figure \ref{Fig5} the calculated LM distributions are compared
to the experimental one \cite{Jep04}. Note, that the theoretical
curves are shifted by 10 MeV/c to the left to be compared to the
experimental data.

The solid line in the figure shows the total LM distribution
obtained with all $^{14}$N states shown in Table \ref{Tab3}.
The dashed line represents the LM distribution from
proton removal leading to $^{14}$N in the ground state. As it
corresponds to the smaller proton separation energy (Table
\ref{Tab2}), the LM distribution is narrower, than that for other
$^{14}$N states. Thus, the FWHM value of the total LM distribution
is larger than that for the $^{14}$N ground state by 13 MeV/c.

The consideration of the $^{14}$N production in the excited states
does not change significantly the LM distribution because each
$^{14}$N state (including the ground state) is characterized by
relatively high proton separation energy, and, hence, has nearly
the same (Table \ref{Tab3}) LM distributions. Thus, the value of
FWHM is weakly sensitive to the weights of the $^{14}$N states and
the $^{14}$N excitation. However, these contributions are
essential in the calculations of the proton removal cross section.

\section{Conclusion \label{secV}}

In this paper, we present calculations of the one-proton removal
cross sections from
 $^{15}$O on a Be target at the energy 56 A
MeV. The proton removal cross sections, the $^{15}$O interaction
cross section, and the longitudinal momentum distribution of the
$^{14}$N fragments  are obtained in the eikonal approximation of
the Glauber model with the NN interaction potential. In the
calculations, the production of the $^{14}$N core fragment in the
ground and excited states is regarded. The calculated FWHM=191
MeV/c of the total LM distribution is very close to the
experimentally measured value of $190 \pm 10$ MeV/c \cite{Jep04}.

The calculated value, 95.9 mb, of the total one-proton removal
cross section is also very close to the experimental value $80 \pm
20$ mb \cite{Jep04}. The break-up cross section is about 11\% of
the $^{15}$O interaction cross section.

Returning to the $^{17}$Ne problem, we see that the contribution
of the proton removal from the $^{15}$O core might be essential.
In particular, at the energy 66 A MeV (see experimental data
\cite{kan03}), we get the cross section of the proton removal from
the core fragment 94.4 mb. Due to the weakly-bound protons
blocking the $^{15}$O core in $^{17}$Ne, this cross section is
reduced, contributing about 51 mb to the total one-proton removal
cross section. The contribution of the valence proton removal in
$^{17}$Ne with the spectacular $^{15}$O core is about 110 mb
\cite{Gri05}. Thus, the calculated total proton removal cross
section will be 161 mb. This value satisfies the experimental one,
$168\pm 17$ mb \cite{kan03}. Note, that the contribution of the
proton removal from the $^{15}$O core affects also the width of
the total $^{15}$O LM distribution.

As a result, in the reactions with $^{17}$Ne, the proton removal
cross section measured at the energy 66 A MeV on a Be target
\cite{kan03} is relatively large compared to the cluster model
($^{15}$O$+p+p$) predictions \cite{Gri05} and the measured
$^{15}$O LM distribution is wider than calculated one.

Therefore, the proton removal from the core should necessarily be
 taken into account in calculations
of the $^{17}$Ne fragmentation.




\end{document}